%% file: ms.tex
\RequirePackage{cmap}
\RequirePackage[T1]{fontenc}
\RequirePackage{fix-cm}
\documentclass[twocolumn]{svjour3}          
\smartqed  
\usepackage{cmap}
\usepackage[T1]{fontenc}

\usepackage{csquotes}

\usepackage{microtype}

\usepackage[short]{optional}

\usepackage{graphicx}

\graphicspath{{graphics/}}

\usepackage[american]{babel}

\usepackage[sort,square,numbers]{natbib}

\usepackage{paralist}

\usepackage{url}
\makeatletter
\g@addto@macro{\UrlBreaks}{\UrlOrds}
\makeatother

\usepackage{hyperref}
\hypersetup{hidelinks,
	colorlinks=true,
	allcolors=black,
	pdfstartview=Fit,
	breaklinks=true}
\usepackage[all]{hypcap}

\usepackage[capitalise,nameinlink]{cleveref}
\usepackage{caption} 

\usepackage[IEEE,
publication={SICS Software-Intensive Cyber-Physical Systems},
doi={10.1007/s00450-019-00411-y},
doiText={https://doi.org/10.1007/s00450-019-00411-y},
nocopyright,
nourl
]{authorarchive}

\begin{document}

\title{Transactional Properties of Permissioned Blockchains}

\author{Ghareeb Falazi        \and
        Vikas Khinchi	\and
        Uwe Breitenb{\"u}cher \and
        Frank Leymann 
}


\institute{G. Falazi, U. Breitenb{\"u}cher and F. Leymann \at
              Institute of Architecture of Application Systems, \\
              University of Stuttgart, Germany \\
              \email{\{falazi,breitenbuecher,leymann\}@iaas.uni-stuttgart.de}           
           \and
           V. Khinchi \at
           Diconium Digital Solutions GmbH, Stuttgart\\
           \email{vikas.khinchi@diconium.com}
}

\date{Received: date / Accepted: date}

\maketitle

\begin{abstract}
Traditional distributed transaction processing (TP) systems, such as replicated databases, faced difficulties in getting wide adoption for scenarios of enterprise integration due to the level of mutual trust required.
Ironically, public blockchains, which promised to solve the problem of mutual trust in collaborative processes, suffer from issues like scalability, probabilistic transaction finality, and lack of data confidentiality.
To tackle these issues, permissioned blockchains were introduced as an alternative approach combining the positives of the two worlds and avoiding their drawbacks.
However, no sufficient analysis has been done to emphasize their actual capabilities regarding TP.
In this paper, we identify a suitable collection of TP criteria to analyze permissioned blockchains and apply them to a prominent set of these systems.
Finally, we compare the derived properties and provide general conclusions.
\end{abstract}

\input{introduction}
\input{motivation}

\input{bcAsTP}
\input{transactionalProperties}
\input{summaryOfResults}

\bibliographystyle{spbasic}
\interlinepenalty=10000
\bibliography{ms}
%

\end{document}

%% file: introduction.tex
\section{Introduction}
\label{sec:introduction}
Distributed TP systems have been around for several decades.
Nonetheless, they failed to gain adoption in cases where different enterprises need to collaborate together because of the level of mutual trust needed when jointly running them.
On the other hand, public blockchains were introduced as a means to facilitate trustless collaborations.
However, their open nature resulted in scalability and privacy problems since all peers need to hold a copy of the shared data and process new transactions.
Furthermore, an expensive means to prevent double-spending attempts is needed rendering the approach unusable for B2B applications.
To solve the limitations of both approaches, permissioned blockchains were introduced as replicated databases that have limited mutual trust requirements.
However, the capabilities of permissioned blockchains as TP systems remain unclear due to the absence of a detailed analysis.

In this paper, we bridge the gap between database and blockchain experts by analyzing permissioned blockchains from a TP standpoint in order to answer the research question: \enquote{\textit{What are the transactional properties of permissioned blockchains?}}.
To this end, we
\begin{inparaenum}[(i)]
\item introduce the hypothesis that permissioned blockchains can be viewed as replicated databases, and
\item identify a set of TP properties suitable for describing them. Moreover,
\item we apply the set to a collection of the most prominent technologies in order to infer their characteristics from a TP standpoint. Finally,
\item we categorize these systems based on the results.
\end{inparaenum}
Thus, this analysis supports enterprises willing to incorporate permissioned blockchains into their interactions

The paper proceeds as follows: in \cref{sec:background_and_motivation}, we give a general background about the related technologies and discuss permissioned blockchains and the reasons behind their inception.
Furthermore, we present the major research question, and show that the related work did not address it sufficiently.
In \cref{sec:blockchains_as_replicated_databases}, we propose to look at permissioned blockchains as replicated databases.
Later, in \cref{sec:transactional_properties}, we apply this vision to a number of prominent permissioned blockchains.
Finally, in \cref{sec:summary}, we summarize our findings and detect areas in which these systems can further evolve.

%% file: motivation.tex
\section{Background, Related Work, and Motivation}
\label{sec:background_and_motivation}
In this section, we give background information about TP systems and blockchain systems.
\opt{long}{To this end, we start by introducing what transactions are and what properties we expect from them, next we give further details on the various kinds of systems that process these transactions.
Later, we visit the domain of blockchains and discuss their properties and drawbacks that motivated their evolution over the last decade.
Finally, we motivate the problem we are focusing on and show that the related research did not address it adequately.}

\subsection{Basics of Transaction Processing (TP)}
\label{sec:basics_of_trasnaction_processing}
In computer systems, a \textit{transaction} in its most abstract form is a finite execution of a program that accesses shared data~\cite[p. 2]{bernstein2009principles}.
Each transaction is a unit-of-work that represents a real world \textit{business transaction} executed via a computer system.
This includes, for example, transferring money between two bank accounts.
The code executed by a transaction is known as a \textit{transaction program} which is grouped with others into a \textit{TP application} that automates entire business activities, such as a banking system.
Moreover, a \textit{TP system} is a computer system capable of hosting and managing transaction programs.
Relational databases are the most prominent example of such systems.
Others include messaging systems, and business process management systems.
A TP system is called distributed when the underlying shared data resides in more than one location.

Haerder et al.~\cite{hearder1983acid} define a set of four properties, abbreviated in the acronym ACID, that every transaction must have:
\begin{inparaenum}[(i)]
	\item \textbf{Atomicity} means that a transaction either executes completely, or not at all. 
	This guarantees that the effects of a successful transaction are completely reflected in the underlying datastore, or that none of these effects are reflected if the transaction fails.
	\opt{long}{Partial application of a transaction's operations is not acceptable since it violates the semantics of the modeled business transaction and invalidates the consistency of the underlying data items.}
	\item \textbf{Consistency} means that if a transaction is applied to a consistent datastore, it must preserve this consistency by transforming the state to a new consistent state.
	A datastore is consistent when it preserves a set of predefined domain-specific business rules.
	\opt{long}{This means that consistency mainly stems from the transaction program not from the TP system.~\cite[p.~13]{bernstein2009principles}.}
	\item \textbf{Isolation} means that a transaction should execute as if it is the only one being run at the same time.
	Nonetheless, to achieve higher throughput, TP systems usually run transactions in parallel\opt{short}{ while guaranteeing that they are isolated and do not affect each other, i.e., \textit{serializable}.}
	\opt{long}{. To maintain isolation in this case, TP systems are required to guarantee that the interleaved execution of transactions is \textit{serializable}, i.e., that it is equivalent to a serial execution of these transactions.}
	\item Finally, \textbf{Durability} means that the effects of a successful transaction, i.e., a \textit{committed transaction}, are durably stored in the underlying datastore\opt{long}{, and} even if the TP system suffers from failure\opt{long}{, these effects are still persisted.
	This usually entails that the effects of a committed transaction are stored in stable storage that can survive (normal) system failures}.
	\opt{long}{Durability can also be thought of as a contract between the TP system and its users which states that once a transaction is committed its effects will not be unilaterally revoked by the system.}
\end{inparaenum}

These properties focus on transactions that run on a multi-user centralized TP system, i.e., the underlying shared data is located on a central site.
However, of high importance for the study of blockchains is the concept of \textit{global (or distributed) transactions} which affect data items in more than one site, i.e., in a distributed TP system.
More specifically, we are interested in transactions that operate on replicated data, which are well studied in the field of \textit{replicated database systems}.

\subsection{Replicated Database Systems}
\label{sec:replicated_databases}
Kemme et al.~\cite{kemme2010replication} give a comprehensive overview of this field by identifying the general properties and trade-offs of replicated databases and their effects on transactions, and also by listing the various transaction correctness criteria that contribute to \textit{strong consistency}.

First, replicated databases can be described based on two parameters: 
\begin{inparaenum}[(i)]
	\item the \textbf{location} in which the execution of transactions is possible.
	This is traditionally divided into either the \textit{primary-copy} case that permits the execution of transactions only on a specific replica, or the \textit{update-anywhere} case in which transactions can be executed on any of the replicas.
	In both cases the effects of the transaction are propagated from the origin of execution to the other replicas according to some protocol.
	This leads to the second parameter, namely
	\item the \textbf{synchronization strategy} between replicas.
	Traditionally, this also has two cases: the \textit{eager strategy}, in which the replica that executes the transaction transmits its effects to the others and ensures they will commit them before signaling success to the user, and the \textit{lazy strategy}, in which the executing replica announces the success of the transaction immediately after its local commitment and before ensuring its success on the other replicas.
	This indicates a clear trade-off between \textit{availability} and \textit{strong consistency}~\cite[p.~261]{bernstein2009principles}.
\end{inparaenum}

To have a better understanding of what we mean by strong consistency, we take a look at three of the correctness dimensions that collectively contribute to this notion~\cite{kemme2010replication}:
\begin{inparaenum}[(i)]
	\item \textbf{Global atomicity} is a generalization of ACID atomicity (c.f. \cref{sec:basics_of_trasnaction_processing}) but for a replicated environment.
	It requires that a global transaction either entirely commits or entirely aborts at \textit{all} replicas. 
	\opt{long}{There must be no replica that commits the transaction while others abort it and vice versa.}
	Traditionally, this is achieved via coordination protocols such as two-phase commit (2PC).
	Strong consistency here requires that all data copies have the same value at commit time, whereas \textit{weak consistency} either refers to the possibility of reading stale data values from certain replicas even when others report a successful commit, i.e., \textit{staleness}, or it refers to temporal \textit{inconsistency} that need to be resolved.
	\item \textbf{Global isolation}, is a generalization of the ACID isolation.
	The global isolation level usually considered is \textit{1-copy-serializability} (1CS) in which the execution of a set of interleaved global transactions is equivalent to a serial execution of these transactions on a single logical copy of the database even if some replicas are failing.
	\opt{long}{Another important global isolation level is \textit{snapshot isolation} in which each of the concurrent transactions is evaluated against a snapshot of the local database at each replica, and at transaction commit time conflicts of write operations only are not allowed even among different replicas.}
	Finally, 
	\item \textbf{session consistency} addresses a sequence of global transactions issued from the same user.
	A consistent session guarantees that transactions are committed in the order they were issued.
	\opt{long}{Specifically, each transaction should sees the effects of previously submitted ones from the same session, i.e., \textit{read-your-writes consistency} across transactions of the same session.}
\end{inparaenum}
We use these dimensions later to categorize permissioned blockchains from a TP standpoint.

Other properties~\cite{kemme2010replication} that differentiate database replication approaches and are important for our discussion here include:
\begin{inparaenum}[(i)]
	\item The \textbf{execution strategy} which refers to a spectrum of mechanisms.
	On one end of the spectrum, we find the case of all replicas executing every statement of every transaction, i.e., \textit{statement, active, or symmetric replication}.
	On other end, a single replica executes the transaction program statements while generating a write-set of changed variable that the other replicas apply to their copies, i.e., \textit{object, passive, or asymmetric replication}.
	\item \textbf{Concurrency control} on the other hand, refers to the mechanism that ensures the absence of undesired conflicts in the execution history of interleaved global transactions, i.e., global isolation.
	These mechanisms are generally divided into pessimistic ones, e.g., \textit{2-phase-locking} (2PL), and optimistic ones, like \textit{multi-version concurrency control} (MVCC).
	Finally, 
	\item two major \textbf{architectural alternatives} can be identified.
	\textit{Kernel-based replication}, in which the replica control mechanism is implemented in the software of each node.
	On the other hand, \textit{middleware-based replication} delegates the task of replica control to a dedicated software layer that coordinates replication transparently.
\end{inparaenum}

Despite their maturity, a drawback that hinders the adoption of such systems for the purpose of enterprise integration is the level of trust needed to run traditional replication control mechanisms~\cite[p.~241]{bernstein2009principles}
Therefore, we deviate our discussion to blockchain systems.

\subsection{Blockchains}
\label{sec:blockchains}
Blockchains refer to systems that maintain an immutable ledger of transactions shared among a set of mutually untrusting parties.
These parties usually have conflicting interests, e.g., \opt{long}{they could share a buyer-seller relationship, or they could be }a consortium of companies competing in the same market\opt{long}{, or even a set of participants in a collaborative development process of a  mission-critical application in which the accountability of actions should be maintained~\cite{Falazi2018_BlockchainCollaborativeDevelopment}}.
Although mutual trust is missing in such cases, collaboration is often necessary and beneficial to all parties. 
However, this leads to the question of who should manage the joint process; a participant managing the interaction is out of question as they could manipulate it in their own favor. 
One way to solve this is the introduction of a trusted third-party, such as a notary service, a governmental agency, or a payment settlement intermediary.
Nonetheless, this introduces raised transaction fees as well as potential managerial and performance hurdles\opt{long}{ such as a single point of failure, complicated system integration and vendor lock-in}, but above all, the existence of such a trusted third party in certain scenarios is by itself challenging, which is the major motivation behind blockchain systems:\opt{long}{ when configured correctly,} they run as a trusted intermediary which is jointly operated by the untrusting parties in a transparent way that does not favor specific participants over the others.
Moreover, they can be categorized\opt{long}{ based on whether they allow anyone to take part in them or only specific users} into permissionless and permissioned.

\subsubsection{Permissionless Blockchain Systems}
\label{sec:permissionless_blockchains}
A permissionless (or public) blockchain, is a peer-to-peer blockchain open for anyone.
The only condition is to run a local instance of the shared protocol that is connected with the other instances via a network, usually the Internet.
This is comparable to other decentralized systems such as BitTorrent~\cite{qui2004bittorrent}.

As an example, Bitcoin~\cite{nakamoto2008bitcoin} is a peer-to-peer payment network without any settlement intermediaries, and it introduces its own cryptocurrency, which is also called bitcoin.
In this system, sending a payment to another peer entails issuing a digitally signed transaction stating, among other things, the identity of the recipient and the amount of cryptocurrency to be sent.
The transaction then propagates through the network using a gossip protocol~\cite{birman2007gossip}.
All peers receiving the transaction validate its integrity by verifying the authenticity of the accompanied digital signature as well as by making sure the sender actually owns the coins being transacted.
To this end, every peer in the network maintains an immutable list of transactions grouped into blocks which represents the history of all valid interactions that ever took place in Bitcoin.
Furthermore, in order for all peers to maintain a single view of this shared data structure, which is known as the blockchain, a consensus mechanism is required.
Bitcoin introduces a mechanism based on Proof-of-Work (PoW)~\cite{dwork2003pow} in which the blockchain advances periodically one block at a time.
To achieve this, certain peers, known as miners, formulate blocks out of the valid transactions they recently received so that every one of them proposes a single new block in each round.
\opt{long}{These blocks are not necessarily identical since their constituent transactions were received via the gossip protocol which is only eventually consistent.}
\opt{long}{However, since}\opt{short}{Since} only one new block can be appended at a time, miners participate in a probabilistic lottery based on PoW to determine the winning block, which propagates afterwards to the other peers using a similar gossip protocol.
The winning miner receives some bitcoins as a reward and a compensation for the spent energy.
Furthermore, the blockchain is arranged in a way that makes it hard to alter old blocks without controlling the majority of the computational power of the whole network, ensuring immutability.

Public blockchains, in general, \opt{long}{share with Bitcoin the following properties:
\begin{inparaenum}[(i)]
	\item they are peer-to-peer systems which means that the various nodes have the same responsibilities, or at least they can freely choose to perform as many responsibilities as the others.
	For example, in Bitcoin, a node can freely choose to be a miner or an ordinary node.
	\item They also manage a common immutable data structure, which chains transactions or groups of them together and advances over time.
	This data structure, which is usually a blockchain~\cite{nakamoto2008bitcoin,wood2018ethereum}, or a DAG~\cite{popov2017tangle} represents the history of events that took place in the system and allows peers to calculate the current world-state and to validate new transactions.
	\item Furthermore, they usually have a mechanism to motivate peers to correctly perform the consensus protocol including the validation of new transactions despite the potential associated cost, e.g., the energy spent to perform PoW.
	In most cases, this mechanism is based on giving rewards using a self-introduced cryptocurrency.
	The real-world value of this cryptocurrency is heavily influenced by the correct functioning of the system.
	This means that peers that own this virtual money are motivated to maintain the system correct and dependable.
	\item Finally, these systems have a mechanism that makes it very difficult to alter the history of events that took place, i.e., the shared data structure.
	Usually this is done by introducing some real-world cost associated with trying to advance the history. 
	In the case of PoW, this is the cost of consumed energy and the required special equipment needed for mining.
\end{inparaenum}

On the other hand, permissionless blockchains }suffer from a number of drawbacks\opt{long}{ that limit their applicability to certain use-cases}:
\begin{inparaenum}[(i)]		
	\item Due to their design, they have a scalability issue in terms of the amount of transactions the network can process.
	For example, Bitcoin sustains only about 7 tx/s~\cite{cormen2016scaling}.
	\item Permissionless blockchain transactions never reach absolute finality.
	Due to the protocol design, there is always a probability that an alternative (partial) history that may include transactions that contradict with some of the currently accepted ones appears in the network.
	If such a history matches a specific criteria, e.g., being the longest chain in the case of Bitcoin, it will be considered as the one true history by the network peers.
	\opt{long}{Although the probability of a block being replaced by another one quickly diminishes when adding newer blocks on top of it~\cite{nakamoto2008bitcoin}, public blockchain transactions never reach a state that guarantees that they will not be invalidated~\cite{falazi2019modeling}.}
	\item Finally, they are public in nature and allow anyone to obtain a unique identity that facilitates issuing new transactions and reading the entire history.
	However, certain applications require more control over the identity of allowed participants as well as the access to shared data.
	For example, a distributed sharing of health care data records among a set of providers should be protected from public access.
\end{inparaenum}
Although certain approaches exist to circumvent some of these drawbacks using, for example, off-chain scaling solutions~\cite{cormen2016scaling}, or cryptography, a need for a scalable, robust blockchain capable of supporting enterprise applications exists in the form of permissioned blockchains.

\subsubsection{Permissioned Blockchains}
\label{sec:permissioned_blockchains}
Permissioned blockchains are systems operated by known entities and facilitate the interactions among them even if they do not have mutual trust.
In contrast to their permissionless counterparts, participation in permissioned blockchains is not open to the public.
This means that they have a mechanism to control the distribution of recognized identities to permitted users, and potentially control what rights identified users have in the system.
For example, certain users may only have the right to read the history of events but not append to it.
Other users may have the right to validate new transactions.
This means that these systems are usually not peer-to-peer since different participants have different roles.
Furthermore, the management of permissions introduces a level of centralization in the system, even though this task is not necessarily done by a single entity: one or more entities become superusers or administrators and need to be trusted and agreed upon by the others.

Nonetheless, controlling participation in the system has the ability to solve the issues we have seen in the case of permissionless blockchains:
\begin{inparaenum}[(i)]
	\item the existence of an access control mechanism solves the problem of privacy; entities not allowed into the system are not permitted to read the shared data or write to it via transactions.
	Furthermore, certain permissioned blockchains have the ability to limit access to specific parts of the shared data even among recognized entities~\cite{androulaki2018fabric, quorum}.
	\item Moreover, having a static group of users that only allows membership changes via explicit reconfiguration allows using the well-studied and well-optimized consensus techniques designed for Byzantine fault-tolerant (BFT) systems~\cite{cachin2017consensus}.
	Depending on network configuration and synchrony assumptions, these techniques are able to achieve a transaction throughput in the order of 80,000 tx/s~\cite{guerraou2010consensus}, which is a huge improvement over the throughput measured in the range of tens of transactions per second current permissionless blockchains can achieve.
	\item Finally, the usage of these consensus protocols allows reaching transaction finality; they usually guarantee reliable broadcast which entails that all honest nodes will reach the same consensus outcome~\cite{cachin2017consensus}.
	This is convenient for external applications that interact with permissioned blockchains since they do not need to implement additional measures to account for the possibility that an accepted transaction gets dropped from the system.
\end{inparaenum}

\subsection{Motivation and Related Work}
\label{sec:motivation_and_related_work}
Permissioned as well as permissionless blockchains do not operate in isolation.
In fact, they can be viewed as software connectors providing communication, coordination and other services for the various software components of large distributed systems operated by different entities~\cite{xu2016connector}.
With this view in mind, we see that it is important to facilitate integrating the usage of blockchains into existing systems.
To this end, studying the transactional properties of blockchains can be very beneficial since it allows experienced software architects to compare them to other data management systems such as relational and NoSQL databases whose behavior is well-understood.
Studying the transactional properties means looking at blockchains as TP systems and drawing conclusions on the guarantees they make regarding how they handle transactions, as well as system-wide properties that affect how other applications can use them as transaction processors.

A previous work by S. Tai et al.~\cite{tai2017salt} analyzed the transactional properties of permissionless blockchains.
They inferred that these systems, by contrast to ACID or BASE systems, are \textit{Symmetric}, since peers have similar roles, \textit{Admin-free}, as they are self governing, \textit{Ledgered}, because a replicated ledger is stored on each node, and \textit{Time-consensual}, since the advancement of the blockchain happens at intervals averaging to a specific targeted value (SALT).
\opt{long}{Moreover, they concluded that the transactions these systems support are \textit{Sequential}, since they are always processed sequentially, \textit{Agreed}, since they go through a consensus process that guarantees their validity, \textit{Ledgered}, as they are always appended to the shared ledger, and \textit{Tamper-resistant}, since they cannot be manipulated or censored (also SALT).}
However, since permissioned blockchains have considerably different characteristics compared to their public counterparts, the aforementioned research cannot be considered to cover them.

C. Cachin and M. Vukolic~\cite{cachin2017consensus} gave an extensive comparison of the various advantages and drawbacks of the consensus protocols commonly used by permissioned blockchains.
Nonetheless, their discussion was limited to these attributes only, and they did not explain their effects on the TP guarantees provided by the corresponding systems.
Therefore, in this work, which we view as necessary for enterprises willing to incorporate permissioned blockchains into their interactions, we try to build upon the previous results to derive the TP properties of permissioned blockchains.

%% file: bcAsTP.tex
\section{Permissioned Blockchains as Replicated Database Systems}
\label{sec:blockchains_as_replicated_databases}
As we saw in \cref{sec:permissioned_blockchains}, permissioned blockchains are considered an adaption of their permissionless counterparts that tries to solve the issues that prevents using them in enterprise scenarios.
However, to answer the research question \enquote{\textit{What are the transactional properties of permissioned blockchains?}}, we present here an alternative point-of-view on the topic.
We propose the hypothesis that \textit{permissioned blockchains are an adaption of replicated databases} that introduces trust guarantees influenced from public blockchains, which eases their applicability to enterprise integration use-cases.

To intuitively justify this vision, we point out the observation that, in general, blockchains can be thought of as TP systems, particularly, \textit{distributed TP systems}, as they host transaction applications in the form of smart contracts.
A smart contract is a set of procedures that collectively manage a collection of data items shared among the users of the system.
Each of these procedures gets executed as a unit-of-work, and thus can be thought of as a transaction program.
This makes the whole blockchain a distributed TP system since it supports hosting these programs in a distributed environment.
Moreover, shared data (global state) in such an environment is fully replicated among the peers which is usually ensured by the consensus mechanism.
This makes blockchain systems with their TP capabilities very close to the aforementioned replicated database systems (c.f. \cref{sec:replicated_databases}).
Note that even blockchains that do not support hosting arbitrary smart contracts can also be thought of as distributed TP systems as they still host a fixed set of transaction programs.
For example, the basic Bitcoin functionality supports a single transaction program that atomically transfers funds from a set of source accounts to a set of target accounts.\footnote{In fact Bitcoin supports a limited scripting language that allows even more sophisticated programs.}

It is worth mentioning in this context that in the literature, the term \textit{blockchain transaction} (BC transaction for short) refers to something different from the transactions we are describing here.
A BC transaction is a request message sent by an end-user to the blockchain system in order to issue the execution of a transaction program.
In fact a BC transaction is a well-defined data structure that normally includes, among other things, the required input arguments for the smart contract procedure to be invoked as well as a cryptographic signature that guarantees the integrity of these passed data items, authenticates their originating user, and prevents malicious participants from impersonating other users.
Although the concepts of a transaction in the context of TP systems and a BC transaction are related since the latter is a trigger for the former, one needs to keep the difference in mind since the discussion about the \enquote{transactional properties} of some system majorly refers to \enquote{actual} transactions rather that request messages.

Nonetheless, BC transactions have a larger role in blockchain systems than just request messages: as we have seen earlier, validated BC transactions have total order among them, and they represent transitions of the system state, which is defined as the overall values assigned to the shared data items managed by the system.
On the other hand, blockchain systems guarantee that smart contract procedures execute in a deterministic fashion, thus a BC transaction has a 1-to-1 mapping to an \enquote{actual} transaction, which actually operates on data items and results in a state transition.
Furthermore, as seen earlier, BC transactions are authenticated data structures and thus are good for proving the validity of the state they cause the system to transition to, at least from a cryptographic point-of-view.
Therefore, blockchain systems store a permanent ordered record of BC transactions.
Such a record is capable of generating the system state by traversing it from its start and invoking each corresponding transaction.
In fact, apart from performance considerations, it is enough for a a blockchain system to only store the ordered list of BC transactions, i.e., the blockchain data structure, without storing the actual shared state at all since it can be (re-)calculated on-demand by the aforementioned traversal.

\begin{table}
	\centering
	\caption{Summary of the analogy between replicated databases and blockchain systems.}
	\label{table:properties}
	\begin{tabular}{|l|l|} 
		\hline
		\textbf{Replicated Databases}  & \textbf{Blockchains}                                                          \\ 
		\hline
		tranasction program             & smart contract function                                                       \\ 
		\hline
		transaction application         & smart contract                                                                \\ 
		\hline
		transaction                     & \begin{tabular}[c]{@{}l@{}}smart contract function \\execution \end{tabular}  \\ 
		\hline
		request message                 & transaction                                                                   \\ 
		\hline
		replicated data                 & \begin{tabular}[c]{@{}l@{}}global state and \\ blockchain data structure \end{tabular} \\
	
		\hline
		replica control                 & consensus mechanism                                                             \\
		\hline
	\end{tabular}
\end{table}

\Cref{table:properties} summarizes the introduced analogy.
What makes permissioned blockchains differ from their permissionless counterparts in this context is their enforcement of data confidentiality, their relatively high transaction processing rate, and their general resilience against blockchain forking which ensures most of the favorable transaction consistency and guarantees already provided by replicated databases (c.f. \cref{sec:replicated_databases}).

%% file: transactionalProperties.tex
\section{Transactional Properties of Permissioned Blockchain Systems}
\label{sec:transactional_properties}
To further validate the hypothesis presented in the previous section, we examine here a set of the most prominent permissioned blockchain systems and evaluate their transactional properties.
For each introduced system, we first give a high-level overview of it and then try to analyze its properties based on the previous observation that blockchain systems can be thought of as replicated databases from a TP point-of-view.
To this end, we use the concepts defined in \cref{sec:replicated_databases}.

\opt{long}{On the other hand, we do not provide our own assessment on the correctness of the corresponding consensus protocols as this would be outside the scope of this paper.
Instead, we rely on independent analyses of these protocols if they are available~\cite{cachin2017consensus,soton2018poa}, or on the claims presented in each system's documentation.
Moreover, we do not evaluate the TP performance e.g, transaction throughput or latency, but rather refer the interested reader to existing permissioned blockchains benchmarking efforts~\cite{thakkar2018fabricbenchmark, dinh2017blockbench}.}
\opt{long}{The analysis in this section considers systems individually.
Later, we provide a comparison of the findings.}
\input{ethereum}
\input{ripple}
\input{chain}
\input{multichain}
\input{fabric}
\input{sawtooth}
\input{tendermint}

%% file: ethereum.tex
\subsection{Ethereum-based Permissioned Blockchains}
\label{sec:ethereum}
Ethereum~\cite{wood2018ethereum} is a prominent permissionless blockchain platform that was the first to enable Turing-complete smart contracts (transaction applications) via the introduction of the Ethereum Virtual Machine (EVM), a special stack-based state machine that prevents indeterministic executions by limiting its instruction set, and also prevents infinite executions via a parameter called \textit{gas} that gets gradually consumed by every executed operation until it is exhausted, or the execution is over.
A high level language, e.g., Solidity is used to write smart contracts which get compiled into bytecode and stored on the blockchain.
Due to its wide adoption and continuous evolution, Ethereum constitutes a promising candidate for a permissioned counterpart.
To this end several Etherum-inspired permissioned blockchains were introduced that inherit many of its TP properties:
They validate transactions at every peer serially according to the order dictated in the containing block, which is determined previously by the consensus outcome.
Furthermore, they have a kernel-based architecture that embeds the replication logic into each peer.
However, they change the consensus protocol from the original PoW to a permissioned one.
Next, we discuss some prominent examples of these systems.

\subsubsection{Aura and Clique (Proof-of-Authority)}
\label{sec:PoA}
Aura~\cite{aura} (implemented in Parity Ethereum), as well as Clique~\cite{szilagyi2017clique} (implemented in Go Ethereum) are alternative realizations of the Proof-of-Authority consensus protocol, which is meant to replace PoW in permissioned settings.
They are supposed to reduce the power consumption of the network and increase the TP performance while maintaining a high level of security and data consistency.
However, an independent research~\cite{soton2018poa} involving these protocols showed that they, under realistic network and adversary assumption, fail to guarantee safety, i.e., that the probability of blockchain forking exists.
This has dire effects on the consistency measures of TP.
First of all, local durability of transactions at each peer is not guaranteed since a previously committed transaction can later disappear from the blockchain, which would also affect histories that include it both system-wide and at a single user level (session).
This means that global isolation and session consistency are also not guaranteed.
Finally, global atomicity is also not guaranteed since if a transaction commits in one fork, it might not commit on the other.
All in all, although these protocols provide high availability, they are not safe enough to be used in real B2B scenarios.

\subsubsection{Quorum}
\label{sec:quorum}
Quorum~\cite{quorum} is another permissioned blockchain system based on Ethereum.
Besides its permissioning capabilities, it differs from Ethereum in two major ways:
First, it supports private transactions, by partitioning the state into a public and a private one.
The public state is fully replicated as in Ethereum, whereas its private counterpart is different for each participant.
\opt{long}{A private transaction specifies which peers are meant to process it.
	To this end the Private Transaction Manager (PTM) at the originating peer strips the sensitive payload from the request message and for each intended peer, encrypts it with their public key and sends it to their PTM via HTTPS.
	What remains publicly visible from such a transaction is a hash of the removed payload, and an indicator that this is a private transaction.
	Regular peers just accept such transactions, so the hash becomes part of the public state, whereas peers participating in the transaction retrieve its decrypted payload from their local PTM which allows them to execute the relevant smart contract and save the results in their local private state, and since this state}\opt{short}{Since private state} is not replicated, we will only discuss the TP properties of public transactions.
Second, it supports a modular consensus mechanism by allowing both a \textit{crash fault tolerant} (CFT) protocol based on Raft~\cite{raft} which is suitable for trusted environments, and a BFT protocol called Istanbul Byzantine Fault Tolerance (IBFT)~\cite{lin2017ibft} designed for more general use-cases where one-third of the nodes is tolerated to behave in a Byzantine way.

No independent evaluation of these protocols is available at the time of writing, so providing that the claimed safety properties are accurate, Quorum achieves very good transaction consistency:
First, with the absence of blockchain forking and assuming correct handling of transaction storage at each peer, local durability is achieved.
Furthermore, peers have the same decisions regarding the validity of transactions since they share the same state, they execute transactions serially according to their total order, and validation rules are deterministic.
This means that both global atomicity and isolation are guaranteed.
Moreover, for each peer a transaction counter is maintained in the blockchain which facilitates ordering transaction requests\opt{long}{ and prevents replaying already executed transactions by malicious peers.
Using this counter, which is also known as the \textit{nonce}, peers can guarantee session consistency.}\opt{short}{ allowing peers to guarantee session consistency.}  
Finally, the synchronization strategy can be categorized as eager since peers only commit state changes locally when they are sure other correct peers will also commit them.

%% file: ripple.tex
\subsection{Ripple's XRP Ledger}
\label{sec:ripple}
RippleNet (\url{https://ripple.com/}) is a platform providing a global payments settlement network.
It is powered by an underlying permissioned blockchain system called XRP Ledger~\cite{chase2018ripple} with a built-in cryptocurrency, XRP.
Unlike other built-in cryptocurrencies of public blockchains, XRP does not play a role in the consensus mechanism, but rather is meant as a pure digital asset that facilitates B2B interactions by being an intermediary currency\opt{long}{, and as a means to prevent malicious clients from flooding the network with fake transaction requests}.
Since it does not support smart contracts, XRP Ledger is not considered a general purpose blockchain, but rather has a fixed set of financial and managerial transaction programs.

In order to infer the TP properties of XRP Ledger, let's take a brief look at its transaction flow:
\begin{inparaenum}[(i)]
	\item A client, who is not necessarily previously known, proposes a transaction to one of the known nodes (\textit{servers} in XRP terminology), then
	\item the server broadcasts it to the other servers via a gossip mechanism.
	\item The network periodically engages in a mutli-round consensus mechanism so that a subset of the servers, known as the \textit{validators} agree on the set of transaction requests to be included in the next block.
	This consensus protocol allows every server to define a flexible set of peer validators, or a \textit{Unique Node List (UNL)}, to be considered trusted, so it only evaluates proposals from them.
	\opt{long}{Specifically, a server trusts that the super-majority of its UNL will not collude to falsify data relayed to the network.
	Furthermore, in order for the consensus to tolerate Byzantine faults and remain safe, a large proportion of a server's UNL need to overlap with the other UNLs.
	The minimum overlap is shown to be 90\% in a recent research~\cite{chase2018ripple}.}
	Assuming certain periods of network synchrony, and after some rounds of interactions, every server sees that a super-majority of the validators in its UNL agree on the set of transaction to be included in the next block.
	\item Afterwards, each server individually constructs a new block (\textit{ledger} in XRP terminology) by ordering the transaction requests in a deterministic canonical order, and executing them sequentially based on the state associated with the previous valid ledger.
	\item Finally, each validators signs and announces a hash of the ledger it calculated.
	Afterwards, every server ensures that the hash of the ledger it calculated matches the signed hashes announced by a super-majority of its UNL which would mean that the new ledger is validated.
\end{inparaenum}

Under the assumption that UNLs have enough overlap among them and that no more Byzantine validators exist that can be tolerated, the XRP Ledger is guaranteed to be safe, i.e., that blockchain forking cannot take place~\cite{cachin2017consensus}.
This entails that local transaction durability is guaranteed.
Furthermore, since every server serially executes the same set of transactions in the same order based on the same initial state, global atomicity and isolation are also guaranteed.
Moreover, like Ethereum, XRP Ledger introduces a mechanism to order requests from the same client via a sequence number.
This ensures session consistency.
In summary, XRP Ledger provides good transaction consistency guarantees, but with strict assumptions and static domain-specific programs.

%% file: chain.tex
\subsection{Chain}
\label{sec:chain}
Chain~\cite{chain} is another permissioned blockchain system in the financial domain which provides more flexibility than Ripple (c.f. \cref{sec:ripple}) by supporting deterministic Turing-complete smart contracts that govern the issuance and management of \textit{assets}.
Like Bitcoin, Chain follows the unsepent transaction outputs (UTXO) model by allowing to move assets from a transaction's \textit{inputs} to its \textit{outputs}.
Each output field of a transaction refers to a \textit{control program} that needs to be \textit{fulfilled} in order to \textit{spend} it.
Therefore, an input field wishing to spend an output field of a previous transaction provides arguments that fulfill its control program.
On the other hand, creating a new type of assets requires an input field to refer to an \textit{issuance program} instead.
Moreover, each block in Chain's blockchain data structure refers to a \textit{consensus program} which the next block should fulfill in order for it to be considered valid.
However, the default implementation of this smart contract is rather simple; it provides the public key of a specific \textit{block generator node} which is allowed to issue new blocks, as well as the public keys of $N$ \textit{signer nodes} that are supposed to confirm it.
Out of these, $M$ signatures are enough to consider the block valid.
This reduces the protocol to a traditional Byzantine quorum system that tolerates $F$ arbitrarily faulty signers when $M=2F+1$ and $N=3F+1$ while guaranteeing safety and liveness~\cite{cachin2017consensus}.
However, a malicious or faulty block generator can halt the consensus mechanism, or censor specific transactions by not including them in any block.

When the aforementioned safety conditions are met, Chain nodes can guarantee the local durability of transactions.
Furthermore, a global order of transactions, seen by all nodes, is ensured since there is a single authority providing it, namely the block generator.
These transactions are serially and deterministically validated and executed by every non-faulty node in the network.
This means that global atomicity and isolation are guaranteed.
However, by default, the block generator orders transactions in every block according to their hash which probably does not correspond to the submission order by each client.
Therefore, session consistency is not maintained by Chain.
All in all, Chain provides TP guarantees that match the majority of other permissioned blockchains, but by default, depends on a single leader for ordering transactions and generating blocks which could introduce security and liveness issues.

%% file: multichain.tex
\subsection{Multichain}
Multichain~\cite{greenspan2015multichain} is a permissioned blockchain platform based on the Bitcoin Core software. 
Nonetheless, it introduces permissioning capabilities based on public-key cryptography, and employs a low resource-demanding mining variant.
\opt{long}{Multichain also supports \textit{multicurrencies} and \textit{messaging} features that can simplify the interoperability with Bitcoin.}
Transactions in Multichain are based on the aforementioned UTXO model (c.f. \cref{sec:chain}), and have a flow similar to public blockchains but with a different \enquote{mining} mechanism:
Block generation happens periodically as a result of a consensus protocol among a predefined list of miners.
Afterwards, all network nodes independently verify the correctness of these blocks and deterministically validate and execute the contained transactions serially according to the global order. 
However, in each block generation step, only a sub-set of miners is allowed to compete.
This is determined with a \textit{diversity} parameter which puts a limit on the number of blocks generated by the same miner within a given window.
This means that more than one miner could have the right to produce blocks at the same time, which could lead to alternative forks of the blockchain being adopted by different nodes~\cite{cachin2017consensus}.
Like Bitcoin, these forks would eventually become consistent with a high probability using the longest chain rule.
However, this probabilistic model of transaction finality badly affects TP correctness guarantees, exactly as was the case with Aura and Clique (c.f. \cref{sec:PoA}).
Specifically, local transaction durability, global atomicity and isolation, as well as session consistency cannot be ensured with Multichain.
In summary, the heavy dependence on Bitcoin's architecture and transaction flow makes Multichain inherit its finality issues which could have been avoided by using a BFT protocol.

%% file: fabric.tex
\subsection{Hyperledger Fabric}
\label{sec:fabric}
Hyperledger Fabric~\cite{androulaki2018fabric} (or Fabric) is a modular and configurable open-source permissioned blockchain platform that runs under the umbrella of the Hyperledger Greenhouse (\url{https://www.hyperledger.org/}) managed by the Linux Foundation.
It supports a distinctive architecture called \textit{execute-order-validate} which separates Fabric from most other permissioned blockchains that usually follow the active replication architecture~\cite{correia2010practical}.
\opt{long}{Execute-order-validate aims at resiliency, flexibility, scalability, and confidentiality~\cite{androulaki2018fabric}.}

\begin{figure*}
	\centering
	\includegraphics[width=0.75\textwidth]{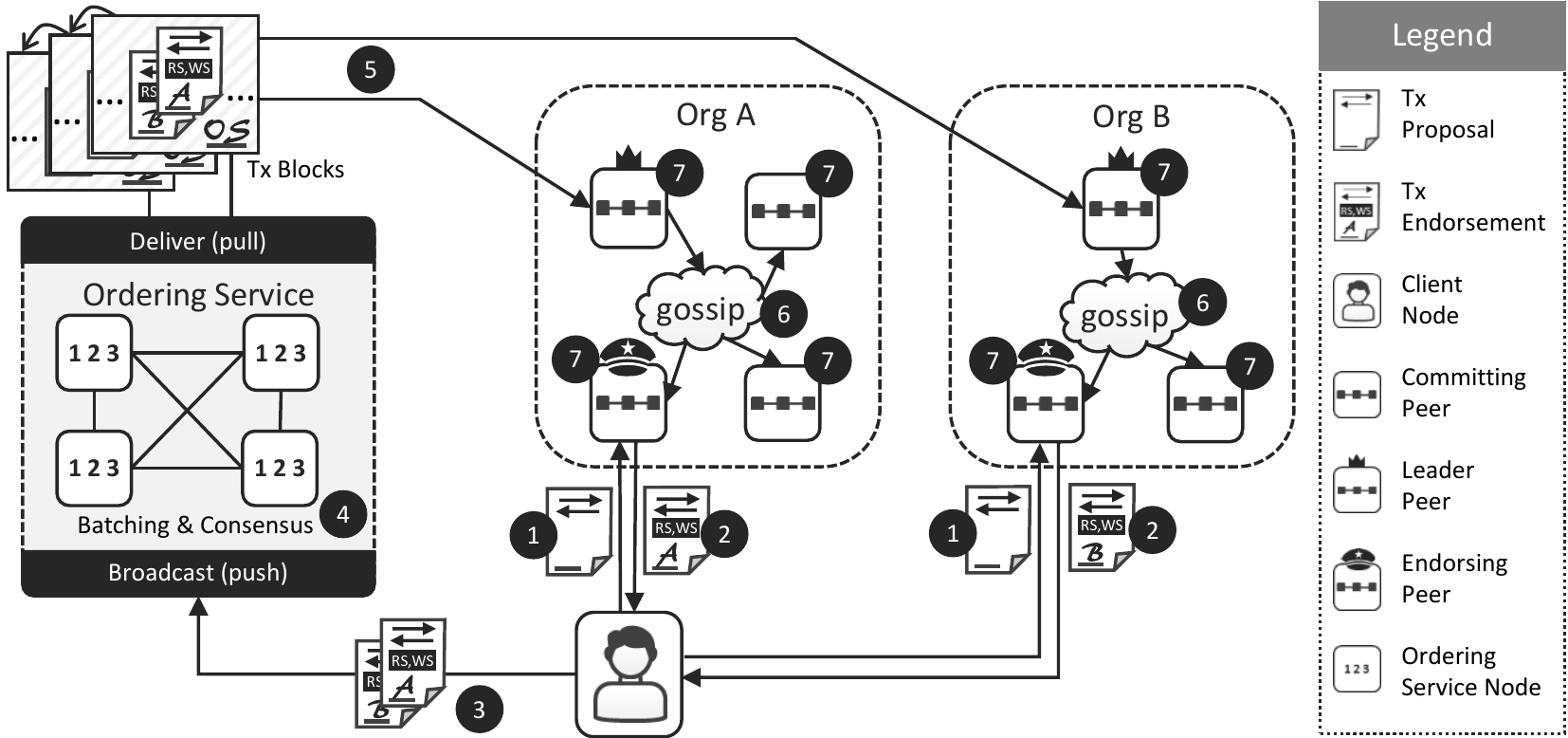}
	\caption{Architecture and transaction flow of Hyperledger Fabric~\cite{androulaki2018fabric}}
	\label{fig:fabric}
\end{figure*}

\Cref{fig:fabric} shows a visual representation of the typical transaction flow in Fabric.
It starts when a pre-registered \textit{client node} formulates a signed \textit{transaction request} (a \textit{request} for short) to execute one function of a smart contract (\textit{chaincode} in Fabric terms).
Fabric introduces a policy-based trust model for the process of endorsing transaction proposals, i.e., verifying their validity and ensuring their conformance to business rules.
To this end, a chaincode-wise \textit{endorsement policy} is defined and stored in the genesis block of the replicated blockchain (\textit{ledger} in Fabric terms).
This policy specifies the set of peers that need to endorse a transaction proposal before it is considered valid.
Such peers are known as \textit{endorsing peers}, and they are usually representatives of the business entities participating in the process.
\opt{long}{An endorsement policy, which models trust relationships among these entities, is defined in a simple domain-specific language that allows monotone logical expressions over the set of endorsers, for example, a policy could state that a set of 2 out of 3 specific endorsing peers should endorse any given transaction proposal.}
The policy applied in \cref{fig:fabric}\opt{short}{, for example,} states that both of the endorsing peers of organization A and B should endorse the given proposal.
Therefore, the client sends it to them (1).
Notice that Fabric divides its peers into organizations that represent the aforementioned real-world business entities.
So within each organization mutual trust is assumed.
An endorsing peer receiving a proposal simulates it by executing the transaction on a stable snapshot of the latest committed state it knows of, and generates a \textit{readset} containing the keys of all accessed data items associated with their versions, and a \textit{writeset} that contains the keys of the changed data items with their proposed new values.
The peer then \opt{long}{does not apply these sets to its local state, but rather signs them (among other entries) and }sends them back to the client (2).
\opt{long}{Chaincode should be deterministic, otherwise the read- and writesets the client collects from different peers could be dissimilar.}
When the client collects enough endorsements according to the policy, it formulates a transaction (BC transaction) out of them and submits it to the \textit{ordering service} via the \textit{broadcast} interface (3).
The ordering service is a collection of special nodes known as the \textit{orderes} which are responsible for batching transactions into blocks that give them total order, and for agreeing on this order via a pre-configured consensus protocol.
Fabric provides multiple options for this.
One implementation with CFT guarantees, is based on Apache Kafka ~\cite{cachin2017consensus}, and a less performant but better resilient implementation is based on BFT-SMaRt which tolerates up to one-third of the orderers behaving in a Byzantine way~\cite{sousa2018bftfabric}.
In both cases, atomic broadcast is guaranteed.
After consensus, transactions are batched into signed blocks and cryptographically chained together to form a blockchain (4).
Apart from the aforementioned \textit{broadcast} interface, the ordering service also exposes a \textit{deliver} interface that permits peers to pull the latest blocks (5).
Optionally, and for scalability reasons, only one peer per organization, namely, the \textit{leader peer}, performs the pulling, and then the block is broadcast via \textit{gossiping} (6).
Finally, when a peer receives a new block it validates it by ensuring that the set of endorsements of each transaction fulfills the endorsement policy and that they include identical read- and writesets.
Afterwards, optimistic concurrency control is performed by ensuring that the versions of the data items included in each readset have not changed since its creation.
At the end (7), the block is persisted in the local ledger, and if the previous validation passed, the \textit{local world-state} is updated by applying the writesets which
happens serially according to their total order.
\opt{long}{
All blocks, including the invalid ones, are included in the ledger to facilitate auditing.
To this end a bitmask that identify the valid blocks is also stored.
}

Fabric gives relatively strong transaction processing guarantees.
It ensures the absence of forking in its ledger, i.e., that all correct peers are guaranteed to see the same transactions and to agree on the outcome of each one of them, due to the implemented concurrency control mechanism (MVCC) and the atomic broadcast property of the ordering service~\cite{cachin2017consensus}.
This means that the durability of transactions is guaranteed locally at each peer, and since Fabric has a recovery mechanism for its failed peers that guarantees the execution of all successful transactions, global atomicity is also achieved.
Moreover, atomic broadcast makes sure that the order of transactions at each peer is the same, and because the execution of writesets is serial, then 1CS is also guaranteed.
On the other hand, a unique property of Fabric is that the execution of the transaction program happens specifically at the peers described in the endorsement policy.
Furthermore, the execution is asymmetric since only some of the peers execute transaction programs, while the others only apply writesets.
Synchronization is considered eager since when a peer applies a writeset, it is sure that all other correct peers will do the same.
However, session consistency is not achieved since stale reads are possible because peers do not coordinate before signaling successful transactions to clients.
\opt{long}{For example, a client might issue a transaction that updates the value of $x$, receives a confirmation that it was committed, and then reads a stale value of $x$ from a peer that has not applied the corresponding writeset yet.}
Finally, clients play a prominent role in the middleware-based replica control mechanism of Fabric, and are expected to maintain state and to be functioning at least during the first phase of the transaction flow.

%% file: sawtooth.tex
\subsection{Hyperledger Sawtooth}
\label{sec:sawtooth_sec}
Hyperledger Sawtooth~\cite{sawtooth} (or Sawtooth), like Fabric, is a general-purpose blockchain platform from the Hyperledger Greenhouse that is both highly modular and configurable.
However, compared to Fabric, Sawtooth focuses more on providing flexible architecture and sophisticated TP capabilities rather than focusing on data confidentiality and privacy.
\opt{long}{Nonetheless, both maintain adequate smart contract expressiveness and permissioning capabilities.}
It introduces \textit{transaction families} as the way to define smart contracts.
Specifically, a transaction family, which can be plugged into the system and can co-exist with other families, describes a predefined set of possible transaction operations that are allowed to be executed.
\opt{short}{One part of every transaction family is the \textit{transaction processor} that implements the needed operations.}
\opt{long}{A custom transaction processor is associated with every family and contains an implementation of the corresponding transaction operations written in one of the many supported high level languages.
This processor communicates with the global state maintained at each peer with a \textit{set and get} interface.
From a TP standpoint, a transaction family is a transaction application with a predefined set of transaction programs.}
On the other hand, transaction requests in Sawtooth are grouped into \textit{batches} by clients.
A batch is an atomic unit-of-work, i.e., either all transactions in a batch are committed, or none of them are, and thus is useful for explicitly defining a set of inter-dependent transactions.
Sawtooth also supports explicit cross-batch dependencies by allowing each transaction to specify a set of other transactions that it depends on.

Transaction flow in Sawtooth follows an \textit{execute-consensus-execute} model.
In this model, \textit{validator} nodes locally execute the batches they receive either directly from clients or from other validators via gossiping.
The execution is based on a snapshot of the last agreed-upon global state which gets incrementally changed when batches are applied to it.
Afterwards, validators organize batches into blocks.
Each block contains in its header the root of a hash tree that describes the world-state after the contained transactions were applied.
Then the configured consensus protocol is executed so that all validators agree on whose version of the next block is to be appended to the blockchain.
Afterwards, validators independently execute the transactions of the agreed-upon block and ensure that the resulting world-state is identical to the one declared in the header.
This guarantees that a badly implemented user-defined transaction family that does not ensure deterministic execution will not cause state inconsistencies.
On the other hand, a prominent feature of Sawtooth is its support for parallel transaction execution.
This is ensured in both of the previously mentioned execution phases via a \textit{parallel scheduler}, which determines, for each transaction, a list of predecessor transactions that need to be executed before it. 
To this end, every submitted transaction proposal should explicitly indicate the sets of state addresses it could \textit{read} from and \textit{write} to.
Using these sets, the scheduler is able to identify conflicting transactions which, combined with the explicit dependency information associated with each transaction, allows it to determine the predecessor lists.
Moreover, if the scheduler detects, at some point in time, that all the predecessors of one or more transactions are already executed, it runs them on parallel allowing for potentially a substantial performance gain.
Sawtooth ensures that this scheduling procedure produces serializable executions and is deterministic across peers, thus ensuring global isolation with 1CS.
As mentioned earlier, the consensus mechanism in Sawtooth is modular and configurable.
This gives a variety of options ranging from PoW-like mechanisms, e.g., Proof of Elapsed Time (PoET), which scales well in terms of node count but do not provide finality, to PBFT-like mechanisms which are not as scalable but guarantee finality even in Byzantine environments.
Assuming that a consensus mechanism that does guarantee finality is used, local transaction durability is ensured due to the absence of forks and maintaining the blockchain data structure and the state on stable storage.
Moreover, all peers are guaranteed to come to the same conclusion on the validity of transactions.
Thus, global atomicity is ensured.
Furthermore, the ability to specifically indicate inter-transaction dependencies allows clients to force session consistency.
Finally, all peers execute the programs of all requested transactions; therefore, the execution strategy is categorized as symmetric.
In summary, from a TP point-of-view, Sawtooth is a general purpose and parallel TP system that allows explicit handling of transaction dependencies and guarantees strong consistency.

%% file: tendermint.tex
\subsection{Tendermint-based Permissioned Blockchains}
\label{sec:tendermint}
Tendermint Core~\cite{buchman2018tendermint} is not a standalone permissioned blockchain, but rather a software middleware that implements a BFT replicated state machine.
A node using Tendermint Core has a local deterministic implementation of a state machine, and by using Tendermint Core, it reaches consensus with the other correct nodes on the set of transactions that will be applied to the state machine and the order in which they will be applied.
To this end Tendermint Core introduces a BFT protocol which is divided into \textit{phases} and \textit{rounds}.
In each phase, the agreement on a new block of ordered transactions is reached using one or more rounds.
Similar to PBFT~\cite{castro2002pbft}, in every round, a leader node proposes a new block to the network and engages in several cycles of authenticated message exchange.
Normally, this will lead to all honest nodes agreeing on the next block.
However, in certain cases, consensus could not be reached which would lead to the initiation of a new round with a new leader.
Nonetheless, under the assumption that the network eventually becomes synchronous, termination is guaranteed while tolerating up to one-third of nodes being Byzantine~\cite{cachin2017consensus}.

As mentioned earlier, Tendermint Core is only a middleware, i.e., other software components rely on it to deterministically, and globally advance the replicated state machine they operate.
\opt{long}{To this end, Tendermint Core provides a callback mechanism, called the Application Blockchain Interface (ABCI), to which other client applications can plug in order to get notified when interesting events occur such as the agreement on the next block.}
Tendermint Core expects these applications to implement the state machine in the way they need to, and to maintain its state locally as well.
It also expects them to reply to its requests to judge on the validity of proposed transactions.
In summary, a client application maintains state, executes transaction programs, and validates transaction requests.
Thus, it is difficult to judge on the TP capabilities of Tendermint Core without considering the client application.
Therefore, we take a look at one specific permissioned blockchain system that uses Tendermint Core as its consensus engine, namely BigchainDB.
BigchainDB~\cite{bigchaindb} is a distributed database that utilizes the blockchain technology to achieve a certain level of decentralization and immutability.
Each BigchainDB node hosts both a Tendermint Core, and a MongoDB (https://www.mongodb.com/) nodes.
It also allows clients to create assets, which represent any physical or digital object, and transfer their ownership to other clients.
From the standpoint of Tendermint Core, BigchainDB is a client application that manages the state of the replicated machine via the MongoDB database, executes a fixed set of deterministic transaction programs that cause the state to transition, and validates transaction proposals against asset double-spending attempts.
With these properties in mind, we can judge on the TP capabilities of BigchainDB.
MongoDB provides stable storage and Tendermint Core guarantees consensus safety, so local durability is guaranteed at each node separately.
Furthermore, Tendermint Core provides all nodes with the same ordered set of transactions which they sequentially and deterministically apply starting from the same initial state.
Thus global atomicity and isolation are guaranteed.
However, session consistency cannot be ensured since a client-wise request sequence number is not maintained.
\opt{long}{This is an issue of BigchainDB rather than Tendermint Core, since the former can include such a mechanism in its TP logic in a way similar to Ethereum for example.}
On the other hand, the replication control of BigchainDB can be categorized as middleware-based due to the dependence on Tendermint Core to achieve consensus and to disseminate transaction blocks.
Moreover, peers do not commit transactions locally until they are sure others will do the same which makes the replication eager.
As a conclusion, we see that judging on the TP characteristics of permissioned blockchains that use Tendermint Core does not only depend on the properties it provides, but also on how these systems implement and manage the associated replicated state machine.

%% file: summaryOfResults.tex
\section{Summary of Results}
\label{sec:summary}

\begin{table*}
	\centering
	\caption{Comparison of the TP correctness criteria and the system-wide properties of the studied systems.}
	\label{table:summary}
	\begin{tabular}{|c|c|c|c|c|c|c|} 
		\hline
		& \textbf{\begin{tabular}[c]{@{}c@{}}Aura, Clique,\\Multichain\end{tabular} }& \textbf{\begin{tabular}[c]{@{}c@{}}Quorum$^a$,\\Ripple\end{tabular}}    & \textbf{Chain$^b$}                                                      & \textbf{Fabric}           & \textbf{Sawtooth$^c$}                                                                & \textbf{\begin{tabular}[c]{@{}c@{}}Tendermint/\\BigchainDB\end{tabular}}  \\ 
		\hline
		Local ACID              & no                                                                & yes                                                        & yes                                                        & yes              & yes                                                                     & yes                                                                \\ 
		\hline
		\begin{tabular}[c]{@{}c@{}}Global\\Atomicity\end{tabular}        & no                                                                & yes                                                        & yes                                                        & yes              & yes                                                                     & yes                                                                \\ 
		\hline
		Global Isolation         & no                                                                & 1CS                                                        & 1CS                                                        & 1CS              & 1CS                                                                     & 1CS                                                                \\ 
		\hline
		\begin{tabular}[c]{@{}c@{}}Session~\\Consistency\end{tabular}     & no                                                                & yes                                                        & no                                                         & no               & yes                                                                     & no$^d$                                                               \\ 
		\hline
		Tx Location     & anywhere                                                          & anywhere                                                   & anywhere                                                   & policy-driven    & anywhere                                                                & anywhere                                                           \\ 
		\hline
		\begin{tabular}[c]{@{}c@{}}Execution~\\Strategy\end{tabular}      & symmetric                                                         & symmetric                                                  & symmetric                                                  & asymmetric       & symmetric                                                               & symmetric                                                          \\ 
		\hline
		\begin{tabular}[c]{@{}c@{}}Synchronization\\Strategy\end{tabular} & lazy                                                              & eager                                                      & eager                                                      & eager            & eager                                                                   & eager                                                              \\ 
		\hline
		\begin{tabular}[c]{@{}c@{}}Concurrency\\Control\end{tabular}      & \begin{tabular}[c]{@{}c@{}}serial~\\execution\end{tabular}        & \begin{tabular}[c]{@{}c@{}}serial~\\execution\end{tabular} & \begin{tabular}[c]{@{}c@{}}serial~\\execution\end{tabular} & MVVC             & \begin{tabular}[c]{@{}c@{}}deterministic\\predecessor list\end{tabular} & \begin{tabular}[c]{@{}c@{}}serial~\\execution\end{tabular}         \\ 
		\hline
		Architecture                                                      & kernel-based                                                      & kernel-based                                               & kernel-based                                               & middleware-based & kernel-based                                                            & kernel-based                                                       \\
		\hline
	\end{tabular}
\\[1.5pt]
\caption*{$^a$Providing that IBFT is used, and considering public transactions only. $^b$Assuming correctly behaving block generator. $^c$Assuming a consensus choice guaranteeing finality. $^d$Other applications using Tendermint can enforce it.}
\end{table*}

In this section, we summarize our findings from \cref{sec:transactional_properties} by providing a comparison of the various permissioned blockchain systems that we considered in terms of TP characteristics.
\Cref{table:summary} shows a summary of the extent to which each considered permissioned blockchain satisfies the TP correctness criteria and what system-wide properties they expose.
We can recognize three categories of systems with regards to the correctness criteria in this table:
\begin{inparaenum}[(i)]
	\item Systems that do not avoid blockchain forking (Clique, Aura, and Multichain), and as a result do not guarantee any of the correctness criteria either.
	These systems evolved from public blockchains, and inherited these properties from them.
	\item Systems that guarantee all of the correctness criteria apart from session consistency (Chain, Fabric and Tendermint).
	\item Systems that guarantee all four of the correctness criteria (Quorum, Ripple, and Sawtooth). 
	However, out of these three systems, only Sawtooth provides general-purpose transaction programs, whereas the other two are domain-specific.
\end{inparaenum}

On the other hand, we quickly recognize that most of the systems expose similar system-wide properties:
\begin{inparaenum}[(i)]
	\item Most systems, apart from Fabric, allow the client to issue the transaction request to any of the peer nodes in the network.
	This peer will then disseminate the request to the rest of the network via gossiping.
	Fabric, on the other hand, has an endorsement policy that indicates to which nodes the client should send the request.
	\item Similarly, in most systems, every peer node executes all transactions, whereas in Fabric, only endorsers execute transaction programs, while the others just apply writesets.
	\item Furthermore, all systems that guarantee the absence of blockchain forks have an eager synchronization strategy meaning that a peer does not commit until it is sure all other correct peers will also commit.
	This is clearly not the case if forks are not avoided.
	\item Moreover, most introduced systems have no need for concurrency control since they execute transactions serially according to a globally agreed-upon order.
	However, this limits the performance of the system since it does not permit parallel executions.
	This is addressed by both Fabric and Sawtooth.
	The former by using a traditional snapshot-based MVCC, and the latter by ensuring that schedulers produce the same predecessor list for the same transaction across the various peers.
	\item Finally, apart from Fabric and Tendermint, all systems implement the replication control logic within the \enquote{kernel} of each network node, whereas Fabric and Tendermint follow an architecture in which a middleware coordinates the transaction flow in the system.
\end{inparaenum}

In summary, we conclude that the safety of the chosen consensus protocol is key to ensuring consistent TP by permissioned blockchains.
Furthermore, we think that there is still a room for these systems to evolve especially in terms of guaranteeing session consistency and supporting parallel transaction processing by incorporating sophisticated concurrency control mechanisms.
Finally, we recognize that this study is not enough to judge on the superiority of one system over the others as it misses important factors such as performance, developer-friendliness, and customer support.
Nonetheless, we see our contributions here as a solid basis that allows software architects to know what to expect from permissioned blockchains and how they could be incorporated into their systems.